\documentclass[a4paper,11pt]{article}

\usepackage{jheppub} 
\usepackage{tikz}
\usepackage{tikz-feynman}
\usepackage{subcaption}
\usepackage{placeins}

\title{Reinterpretation of ATLAS and CMS searches in monojet and mono-\texorpdfstring{$V$}{V} final states:\\
prospects of limits on excited neutrinos}

\author[a]{Gabriela Karkosova Martinovicova,}
\author[a]{Vojtech Pleskot}

\affiliation[a]{Faculty of Mathematics and Physics, Charles University, Prague, Czech Republic}

\emailAdd{gabriela.martinovicova@matfyz.cuni.cz}
\emailAdd{vojtech.pleskot@matfyz.cuni.cz}

\abstract{
    \begin{abstract}
  
  Searches for final states with large missing transverse momentum recoiling against a jet or a hadronically decaying vector boson provide strong constraints on a wide class of physics scenarios beyond the Standard Model.
  In this work, we reinterpret existing ATLAS and CMS monojet and mono-$V$ searches at $\sqrt{s} = 13~\mathrm{TeV}$ in the context of excited-neutrino production.
  Published signal-region selections, post-fit background estimates, and observed event yields are used with simulated excited-neutrino signals to derive upper limits on the production cross-section as a function of the excited-neutrino mass.
  The monojet searches allow excited-neutrino masses of up to approximately 4~TeV to be excluded for representative benchmark scenarios.
  Mono-$V$ searches provide constraints on the parameter space region with large couplings to the SM electroweak gauge bosons and low excited-neutrino masses compared to the compositeness scale.
  This is complementary to the region probed by the monojet searches.
\end{abstract}

    }
    

\usepackage{exc_nu_paper}

\begin{document}
\maketitle
\flushbottom

\section{Introduction}
\label{sec:intro}

The search for physics beyond the Standard Model (SM) is one of the key objectives of the Large Hadron Collider (LHC).
Final states featuring large missing transverse momentum, \ptmiss, recoiling against a single energetic jet or vector boson, commonly referred to as monojet or mono-$V$ signatures, provide useful probes of new physics at hadron colliders~\cite{EXOT-2018-06,CMS-EXO-20-004}.
Such signatures arise naturally in a broad class of models,
including supersymmetry~\cite{Golfand:1971iw}, large extra spatial dimensions~\cite{Arkani-Hamed:1998jmv},
scenarios with weakly interacting massive particles~\cite{Steigman:1984ac}, axion-like particles~\cite{ALPs}, and others.
The ATLAS and CMS collaborations have performed searches for these topologies, placing constraints on many classes of new-physics models.

The possibility that quarks and leptons might not be elementary particles but instead composite objects built from more fundamental constituents (preons) has been discussed in the literature~\cite{PhysRevD.22.184,HARARI197983,FRITZSCH1981319}.
In compositeness models, excited fermions (\fexc) emerge as phenomenological predictions of the underlying substructure~\cite{PhysRevD.42.815}.
Searches for \fexc have been conducted at electron--positron, electron--proton, and hadron colliders over several decades.
At LEP, these searches were performed in electron--positron collisions at centre-of-mass (\cme) energies up to 209~GeV~\cite{L32003,OPAL2002,DELPHI2006,DELPHI1999,ALEPH1998}.
The HERA experiments searched for \fexc in electron--proton collisions at \cme up to 319~GeV~\cite{Zeus:200232,Aaron_2008,Aaron_2009}.
CDF and D0 experiments at the Tevatron conducted searches for excited quarks (\qexc) and excited charged leptons (\lexc) in proton--antiproton collisions at $\cme = 1.8$~TeV and $\cme = 1.96$~TeV~\cite{Toback:2014tea}.
At the LHC, both ATLAS and CMS have searched for \qexc~\cite{EXOT-2014-13,EXOT-2016-21,EXOT-2016-33,EXOT-2013-11,TOPQ-2012-09,EXOT-2015-02,CMS-B2G-12-014,CMS-EXO-13-003, CMS-B2G-14-005,CMS-EXO-17-002,CMS-EXO-23-004} and \lexc~\cite{EXOT-2011-23,EXOT-2012-21,EXOT-2015-01,EXOT-2017-22,EXOT-2020-18,CMS-EXO-10-016,CMS-EXO-11-034,CMS-EXO-14-015,CMS-EXO-18-004,CMS-EXO-22-007} in proton--proton (\pp) collisions at $\cme = 7$, 8, and 13~TeV.
Prospects for excited neutrino (\nuexc), and \lexc searches at future lepton--hadron colliders have also been explored~\cite{Caliskan_2017,Caliskan_2018}.
However, dedicated searches for \nuexc have been scarce.
The most recent limits on \nuexc were set by the ATLAS Collaboration in 2012 using 8~TeV \pp collision data, excluding masses below 1.6~TeV~\cite{ATLAS:2014vih}.
No comparable constraints have been reported using the significantly larger Run~2 datasets collected at $\sqrt{s} = 13~\mathrm{TeV}$, leaving the high-mass regime unexplored.
This motivates the reinterpretation of existing searches to assess their sensitivity to \nuexc production.

In this work, we reinterpret selected ATLAS and CMS monojet and mono-$V$ searches by using the published event selections, event yields and background expectations, and simulating the \nuexc production.
We derive 95\% confidence-level (CL) upper limits on the production cross-section of \nuexc as a function of their mass for benchmark assumptions governing the compositeness parameters.
Although the systematic uncertainty treatment and \nuexc signal simulation are less rigorous than in typical searches by the experimental collaborations, our results demonstrate the potential of ATLAS and CMS mono-object searches to largely extend the excluded part of the \nuexc parameter space.

This paper is structured as follows.
Section~\ref{sec:theory} introduces the theoretical framework describing \nuexc, including their production and decay mechanisms.
Section~\ref{sec:simulation} details the simulation of signal events and detector response.
Sections~\ref{sec:reinterpretation_atlas}, and \ref{sec:reinterpretation_cms} describe the reinterpretation of the ATLAS monojet, and CMS monojet and mono-$V$ searches.
Conclusions are summarized in Section~\ref{sec:conclusion}.
\section{Theoretical Framework}
\label{sec:theory}

The phenomenology of \fexc is commonly described using the effective model introduced by Baur, Spira, and Zerwas (BSZ)~\cite{PhysRevD.42.815}.
The model assumes that \fexc have the same spin as their SM counterparts but possess higher masses, which are not generated by the Higgs mechanism.
In the BSZ approach, both left- and right-handed \fexc components form SU(2) doublets.
The \fexc doublets are represented as
\begin{equation*}
  F^\ast_{L/R} = \binom{\left(\nu_e^\ast\right)_{L/R}}{e^\ast_{L/R}}, \binom{u^\ast_{L/R}}{d^\ast_{L/R}}, \dots,
\end{equation*}
where $\nu_e^\ast, e^\ast, u^\ast, d^\ast$ stand for the excited electron neutrino, electron, up quark, and down quark, respectively, the subscripts $L$, $R$ denote the chirality, and the ellipsis indicates the doublets of the second and third \fexc generations.

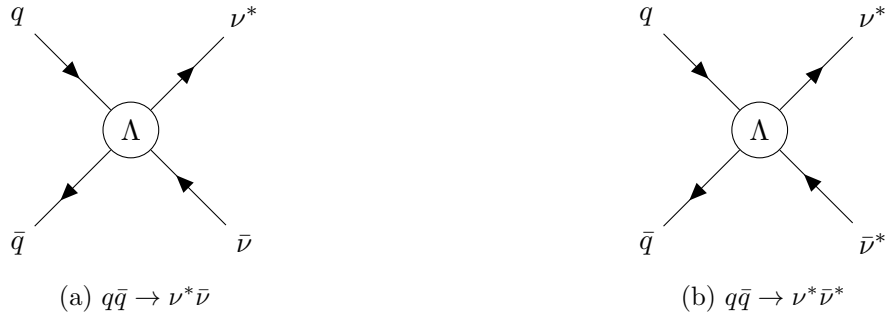
\begin{figure}
  \centering

  \begin{subfigure}{0.45\textwidth}
      \centering
      \begin{tikzpicture}
          \begin{feynman}
              \vertex (i1) at (-1.5, 1.5) {\( q \)};
              \vertex (i2) at (-1.5,-1.5) {\( \bar{q} \)};
              \vertex (v1) at (0,0) [circle, draw] {\(\Lambda\)};
              \vertex (o1) at (1.5, 1.5) {\( \nuexc \)};
              \vertex (o2) at (1.5,-1.5) {\( \bar{\nu} \)};
              
              \diagram* {
                  (i1) -- [fermion] (v1) -- [fermion] (o1),
                  (i2) -- [anti fermion] (v1) -- [anti fermion] (o2),
              };
          \end{feynman}
      \end{tikzpicture}
      \caption{\( q\bar{q} \to \nuexc \bar{\nu} \)}
      \label{fig:single}
  \end{subfigure}
  \hfill
  \begin{subfigure}{0.45\textwidth}
      \centering
      \begin{tikzpicture}
          \begin{feynman}
              \vertex (i1) at (-1.5, 1.5) {\( q \)};
              \vertex (i2) at (-1.5,-1.5) {\( \bar{q} \)};
              \vertex (v1) at (0,0) [circle, draw] {\(\Lambda\)};
              \vertex (o1) at (1.5, 1.5) {\( \nuexc \)};
              \vertex (o2) at (1.5,-1.5) {\( \bar{\nu}^* \)};
              
              \diagram* {
                  (i1) -- [fermion] (v1) -- [fermion] (o1),
                  (i2) -- [anti fermion] (v1) -- [anti fermion] (o2),
              };
          \end{feynman}
      \end{tikzpicture}
      \caption{\( q\bar{q} \to \nuexc \bar{\nu}^* \)}
      \label{fig:double}
  \end{subfigure}

  \caption{Feynman diagrams for \nuexc production via CI.}
  \label{fig:nuexc_production}
\end{figure}

Transitions between excited and ordinary fermions are enabled by two classes of effective operators in the BSZ model.
The first consists of four-fermion contact interactions (CI), which correspond to a low-energy limit of new interactions among the preons:
\begin{equation}
\mathcal{L}_{\mathrm{CI}}
=
\frac{2\pi}{\Lambda^{2}}
j_\mu j^\mu ,
\end{equation}
where $\Lambda$ is the compositeness scale and
\begin{equation}
j_\mu
=
\bar{F}_L \gamma_\mu F_L
+
\bar{F}^\ast_L \gamma_\mu F^\ast_L
+
\bar{F}^\ast_L \gamma_\mu F_L
+ \mathrm{h.c.}
\end{equation}
In the fermionic current $j_\mu$, $F_L$ stands for an SM doublet of fermions with respect to the $SU(2)$ group.
At the LHC, \nuexc are produced via the CI.
Two example Feynman diagrams are shown in Fig.~\ref{fig:nuexc_production}.
The second class of operators consists of gauge-mediated interactions (GI).
The corresponding Lagrangian, relevant for the \nuexc decays, is given by
\begin{equation}
\mathcal{L}_{\mathrm{GI}}
=
\frac{1}{2\Lambda}
\bar{F}^\ast_{R}
\sigma^{\mu\nu}
\left(
g f \frac{\tau^a}{2} W^{a}_{\mu\nu}
+
g' f' \frac{Y}{2} B_{\mu\nu}
\right)
F_{L}
+ \mathrm{h.c.},
\end{equation}
where $f$ and $f'$ are parameters reflecting the composite dynamics.

The \nuexc particles decay through both CI and GI channels.
CI-mediated decays, $\nuexc \rightarrow \nu q \bar{q}$, result in jets and \ptmiss in the final state.
The GI decay modes are
\begin{equation}
\nuexc \rightarrow \nu \gamma,
\qquad
\nuexc \rightarrow \nu Z,
\qquad
\nuexc \rightarrow \ell W.
\end{equation}
When the $W$ or $Z$ boson decays hadronically, the corresponding channel yields one or more jets in the final state.
Commonly, \ptmiss arises from a neutrino that is either produced in the process depicted in Fig.~\ref{fig:single} or from the GI decay to $\nu Z$.
Given these production and decay mechanisms, an important part of \nuexc events results in final states, which are targeted in ATLAS and CMS monojet and mono-$V$ searches.

\section{Signal Generation and Detector Simulation}
\label{sec:simulation}

Monte Carlo (MC) simulation is used to model the signal \nuexc production processes considered in this analysis.
All signal samples are generated for \pp collisions at $\sqrt{s} = 13$~TeV, corresponding to the Run~2 operating conditions of the LHC.
The production and decay of \nuexc are simulated using \textsc{Pythia}~8.310 \cite{bierlich2022comprehensiveguidephysicsusage}, employing the BSZ model to describe both CI production mechanisms and CI and GI decays.
Only the single production of \nuexc is considered, as the \nuexc pair production cross-section is much smaller than the single production one~\cite{PhysRevD.42.815}.
Signal samples are generated for a discrete set of \nuexc masses covering the range of $\mnuexc = $ 600 GeV--5000 GeV.
The simulated mass points are chosen such that interpolation of the exclusion limits is possible.
They are separated by 100~GeV in the regions 600--3000~GeV and 4400--5000~GeV, and by 50~GeV in the region 3000--4400~GeV.
For each mass point, signal samples are generated for several choices of the $f$, $f'$ parameters.
Values $f = f' \in \{0.0, 0.2, 0.4, 0.6, 0.8, 1.0\}$ are used.
Two benchmark settings of the compositeness scale are considered, $\Lambda = \mnuexc$ and $\Lambda = 10~\mathrm{TeV}$.

Detector effects are modelled using \textsc{Delphes}~3.5.0 \cite{DELPHES}.
Two separate detector configurations are employed to approximate the response of the ATLAS and CMS detectors.
They are encoded in detector cards, which are part of the \textsc{Delphes} installation.
In both configurations, jets are reconstructed using the anti-$k_t$ algorithm~\cite{Cacciari:2008gp} with a radius parameter of $R = 0.4$ (AK4).
The threshold for the AK4 jet reconstruction is set to $\pt = 20$~GeV.
A second jet collection is added to the CMS-like configuration, where jets are reconstructed using a larger radius parameter of $R = 0.8$ (AK8).
It emulates the CMS reconstruction of boosted hadronic final states~\cite{CMS-EXO-20-004}.
All simulated events are stored in ROOT~\cite{Brun:1997} format and processed using a custom analysis framework implementing the event selections of the corresponding experimental analyses.

\begin{figure}
\centering
\begin{tabular}{cc}
\includegraphics[width=0.4\textwidth]{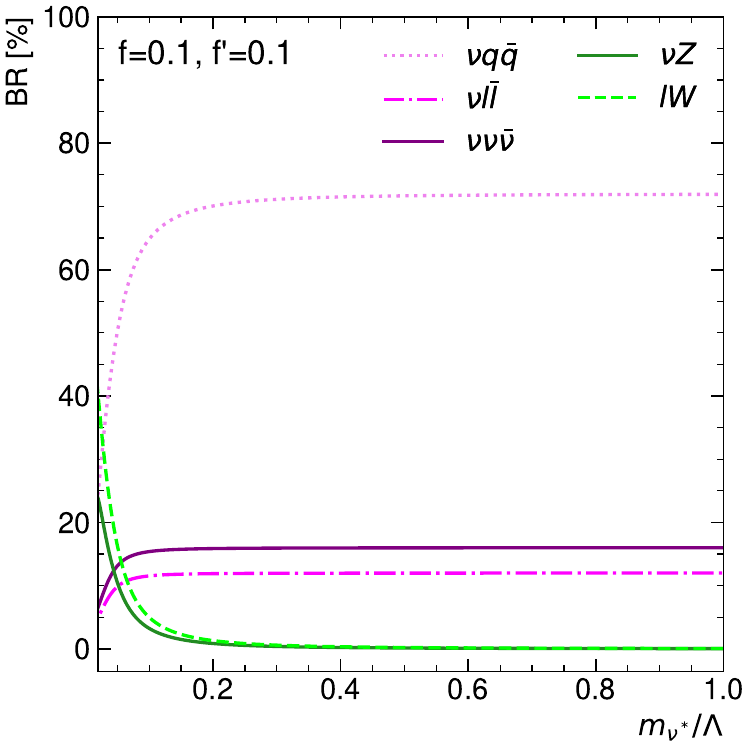} &
\includegraphics[width=0.4\textwidth]{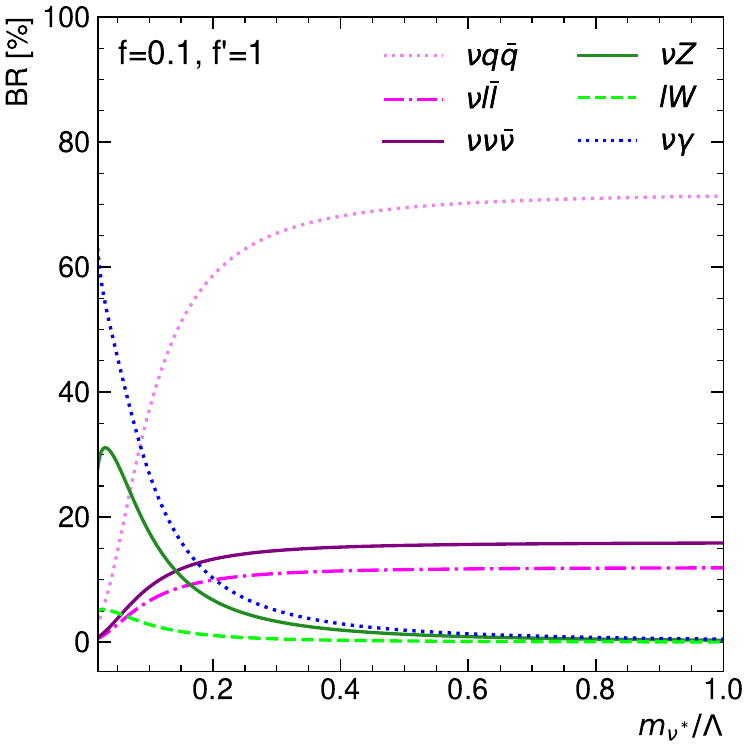} \\[-2mm]
(a) $f=0.1,\;f'=0.1$ & (b) $f=0.1,\;f'=1$ \\[3mm]
\includegraphics[width=0.4\textwidth]{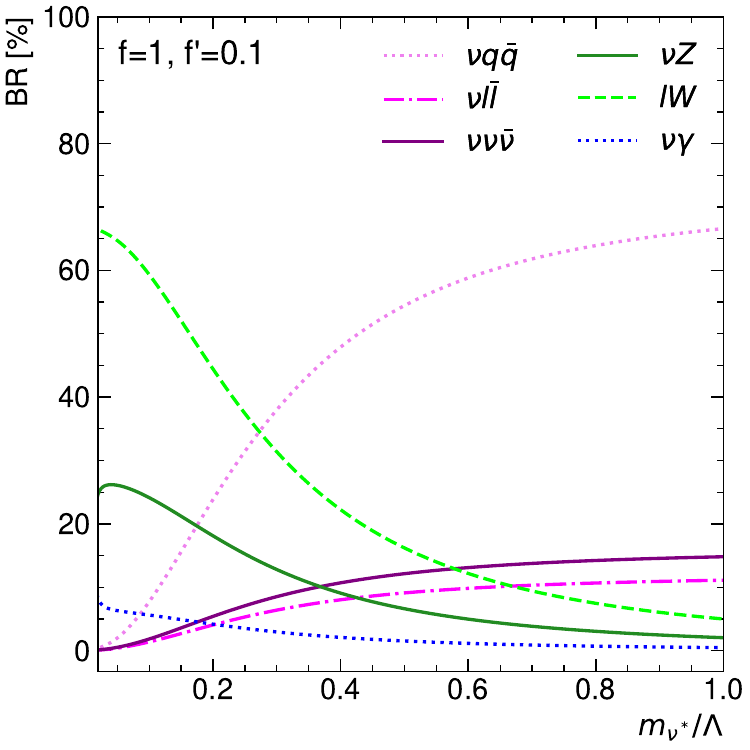} &
\includegraphics[width=0.4\textwidth]{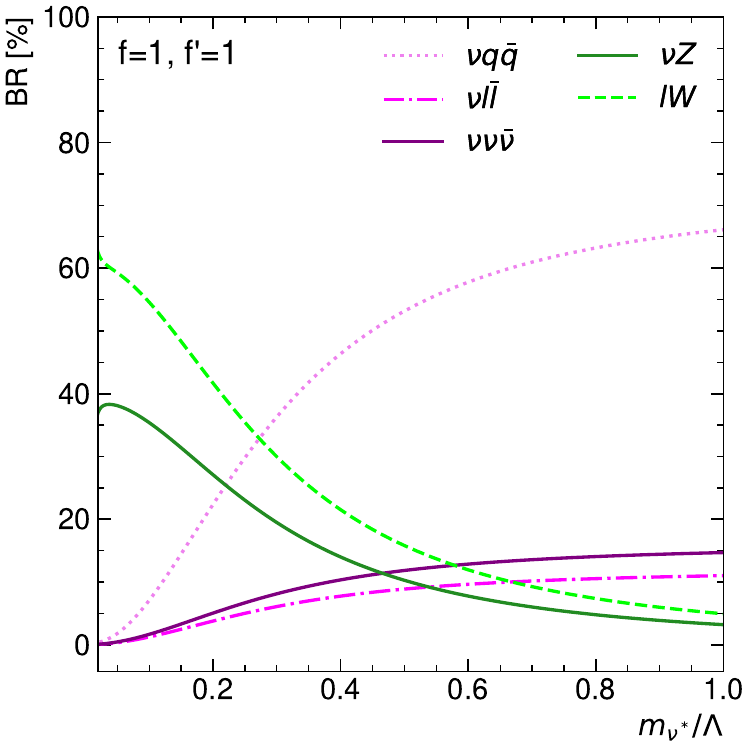} \\[-2mm]
(c) $f=1,\;f'=0.1$ & (d) $f=1,\;f'=1$
\end{tabular}
\caption{
BRs of \nuexc decays as a function of $m_{\nu^\ast}/\Lambda$ for $\Lambda = 10$~TeV and four configurations of the $f$, $f'$ values.
}
\label{fig:BR_grid}
\end{figure}

The \nuexc branching ratios (BRs) are evaluated to learn in which phase space regions one can expect the signal.
The BRs are calculated with \textsc{Pythia} and shown in Figure~\ref{fig:BR_grid} as a function of $\mnuexc/\Lambda$ for $\Lambda = 10$~TeV and four configurations of the $f$, $f'$ values: $(f, f') = (0.1, 0.1)$, $(0.1, 1)$, $(1, 0.1)$, and $(1, 1)$.
The CI decays, $\nuexc \rightarrow \nu q \bar{q}$, $\nu \ell\bar{\ell}$, $\nu \nu\bar{\nu}$, dominate for high values of $\mnuexc/\Lambda$ or low values of the $f$, $f'$ parameters.
The $\nu q \bar{q}$ channel largely dominates the CI decays.
The GI decays, $\nuexc \rightarrow \nu Z$, and $\ell W$, become significant for low values of $\mnuexc/\Lambda$ and high values of the $f$, $f'$ parameters.
The BR of the $\ell W$ channel increases with $f$.
The $\nu Z$ and $\nu \gamma$ channels are the dominant GI decays when $f'$ is large, and $f$ is small.
The $\nu \gamma$ decay is forbidden when $f = f'$.
\section{Reinterpretation of the ATLAS monojet search}
\label{sec:reinterpretation_atlas}

The ATLAS search presented in Ref.~\cite{EXOT-2018-06} is based on events with a high-\pt jet and large magnitude of the missing transverse momentum, \met.
It uses the full 13~TeV \pp Run~2 dataset collected by the ATLAS detector~\cite{refId0} corresponding to an integrated luminosity of 139~fb$^{-1}$.
Events are required to contain at least one AK4 jet with $\pt > 150$~GeV and $|\eta| < 2.4$, and no more than three additional AK4 jets with $\pt > 30$~GeV and $|\eta| < 2.8$.
In the signal region (SR), \met is required to be larger than 200~GeV.
Additional SR selection criteria, like veto of electrons, muons, photons, or presence of hadronically decaying tau leptons in the event, quality requirements on the leading-\pt jet reconstruction, and angular separation between \ptmiss and each jet, are described in Ref.~\cite{EXOT-2018-06}.

The dominant SM background arises from $Z(\nu\nu)$+jets production, with additional contributions from $W(\ell\nu)$+jets, top-quark, and diboson processes.
Dedicated control regions (CRs) enriched in $Z(\ell\ell)$+jets, $W(\ell\nu)$+jets, and \ttbar events are used to improve the background prediction, which is mostly based on MC simulation.
Each region is binned in the magnitude of the \ptrec vector, \ptrecmag; \ptrec is a proxy for the transverse momentum of the system which recoils against the hadronic activity in the event.
In the SR, \ptrecmag is equal to \met.
The ATLAS analysis provides an SR background estimate determined in a simultaneous background-only profile likelihood fit to the CRs together with its total uncertainty and the total data yield in each bin~\cite{hepdata.102093}.

The present study uses the profile likelihood ratio hypothesis test~\cite{Cowan:2010js} and the CL$_\mathrm{s}$ method~\cite{Junk:1999kv,Read:2002hq} to derive exclusion limits on the \nuexc model parameters.
The study uses the asymptotic distributions~\cite{Cowan:2010js} of the test statistic under the signal-plus-background and background-only hypotheses to evaluate the corresponding $p$-values needed in the CL$_\mathrm{s}$ method.
All SR bins defined in Ref.~\cite{EXOT-2018-06} are used in the test.
The background estimate provided by the ATLAS analysis is used as the background template in the likelihood construction.
For each setting of the $\Lambda$, \mnuexc, $f$, $f'$ values, the signal template is obtained by applying the SR selection to the corresponding simulated signal sample.
The event yield in each bin is treated as a Poisson-distributed variable, and the likelihood includes the product of the Poisson probabilities across all SR bins.
The expectation value of each Poisson term is the sum of the signal and background contributions in the corresponding bin.
The signal contribution is scaled by a signal strength parameter, which is the parameter of interest in the test.
The likelihood also includes the total background uncertainty in the form of one Gaussian-constrained nuisance parameter.
The total uncertainty is treated as correlated across all SR bins.
The likelihood functions are built, and all statistical tests are performed using the \textsc{HistFitter} package~\cite{Baak:2014wma}.

\begin{figure}
\centering
\begin{subfigure}[b]{0.59\textwidth}
    \centering
    \includegraphics[width=\textwidth]{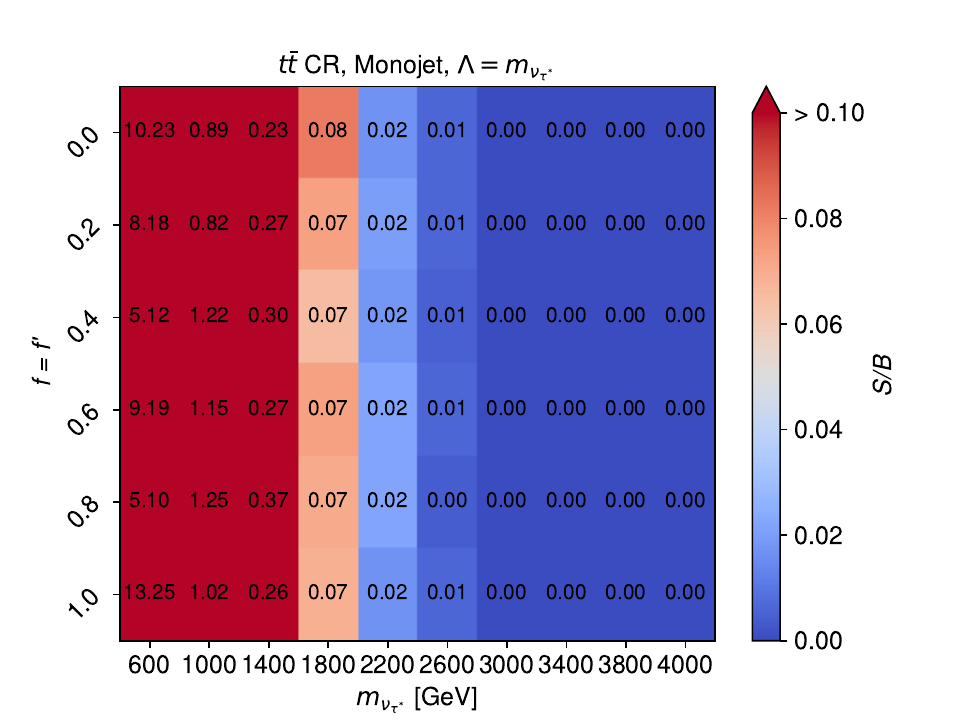}
    \caption{}
    \label{fig:atlas_sig_cont}
\end{subfigure}
\begin{subfigure}[b]{0.40\textwidth}
    \centering    
    \includegraphics[width=\textwidth]{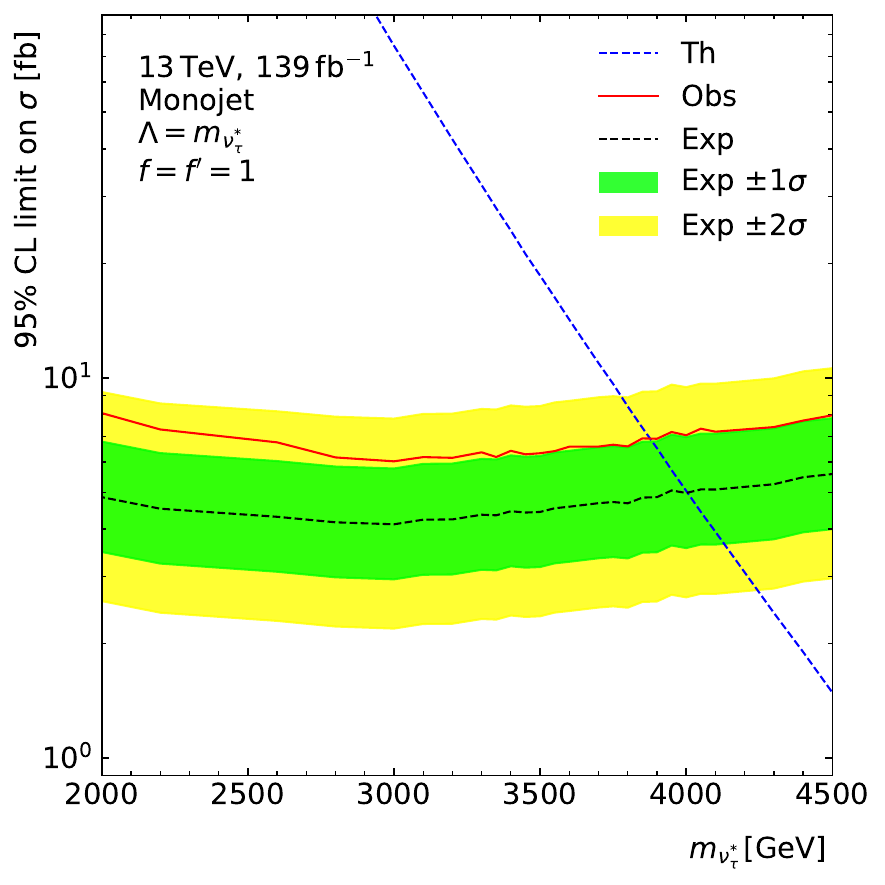}
    \caption{}
    \label{fig:atlas_xs_lim}
\end{subfigure}
\caption{
(a) The \sbr of \nuexct events in the high-\ptrecmag subregion of the ATLAS \ttbar CR as a function of \mnuexct and $f$.
The \ptrecmag threshold is 600~GeV.
The $\Lambda$ parameter is set equal to \mnuexct, and $f' = f$.
The dark red colour highlights $(\mnuexct, f)$ points with signal yield higher than 10\% of the post-fit background estimate.
(b) The 95\% CL upper limit on the \nuexct production cross-section as a function of \mnuexct for $\Lambda = \mnuexct$, and $f = f' = 1$.
The observed (expected) limit is shown in a solid red (dashed black) line.
Boundaries of the green (yellow) band display $\pm 1\sigma$ ($\pm 2\sigma$) uncertainty in the expected limit.
The dashed blue line displays the theoretical cross-section.
The limits are based on the ATLAS monojet search results.
}
\label{fig:atlas_sig_cont_xs_lim}
\end{figure}

A key aspect of the reinterpretation is ensuring its logical consistency.
Since the ATLAS analysis relies on CRs to normalize and shape the background prediction, a significant signal contamination in these regions could invalidate the post-fit background estimate.
If the signal existed and was present in the CRs, it would bias the background prediction in the CRs and the SR.
To address this, the \nuexc signal yields are computed for all ATLAS CRs: $W(\mu\nu)$, $W(e\nu)$, $Z(\mu\mu)$, $Z(ee)$, and \ttbar.
As the signal is expected to be most prominent in the high-\ptrecmag part of each CR, the signal and background yields are compared there.
In each CR, a subregion is defined with a requirement on \ptrecmag to be above a certain threshold, \ptrecth.
The threshold is chosen such that the signal yield in that subregion is non-zero, with a statistical significance of at least $2\sigma$, for $\mnuexc = 1400$~GeV, $\Lambda = \mnuexc$, and any value of $f = f'$.
The value $\ptrecth = 600$~GeV satisfies this requirement for all CRs and \nuexc flavours.
If the signal yield exceeds 10\% of the background in the high-\ptrecmag subregion of any CR, the corresponding point in the \nuexc model parameter space is deemed unreliable for reinterpretation.
When the $\Lambda$ parameter is set to 10~TeV, and $f = f'$, the signal contamination is negligible in all CRs for all values of \mnuexc between 600~GeV and 4000~GeV (steps of 400~GeV are made), all values of $f$ between 0 and 1 (steps of 0.2 are made), and all \nuexc flavours.
The excited tau neutrino (\nuexct) signal contamination exceeds the 10\% threshold when $\Lambda = \mnuexct$, and $\mnuexct \leq 1.4$~TeV.
With this setting, the \ttbar CR turns out to have the highest relative \nuexct signal contamination.  
Figure~\ref{fig:atlas_sig_cont} displays the \nuexct signal-to-background ratio (\sbr) in its high-\ptrecmag subregion.
Masses $\mnuexct \leq 1.4~\mathrm{TeV}$ are excluded from the subsequent statistical test of the $\Lambda = \mnuexct$ scenario for all values of $f = f'$.
The same constraints are obtained for the \nuexce signal.
For the excited muon neutrino (\nuexcm) signal, the signal contamination is the most prominent in the $W(\mu\nu)$ CR.
Parameter space points with $\mnuexcm \leq 1~\mathrm{TeV}$ ($\mnuexcm \leq 1.4~\mathrm{TeV}$) are excluded from the statistical test of the $\Lambda = \mnuexcm$ scenario for all values of $f = f'$ (for $f = f' \geq 0.4$).

\begin{figure}
\centering
\begin{subfigure}[b]{0.45\textwidth}
    \centering
    \includegraphics[width=\textwidth]{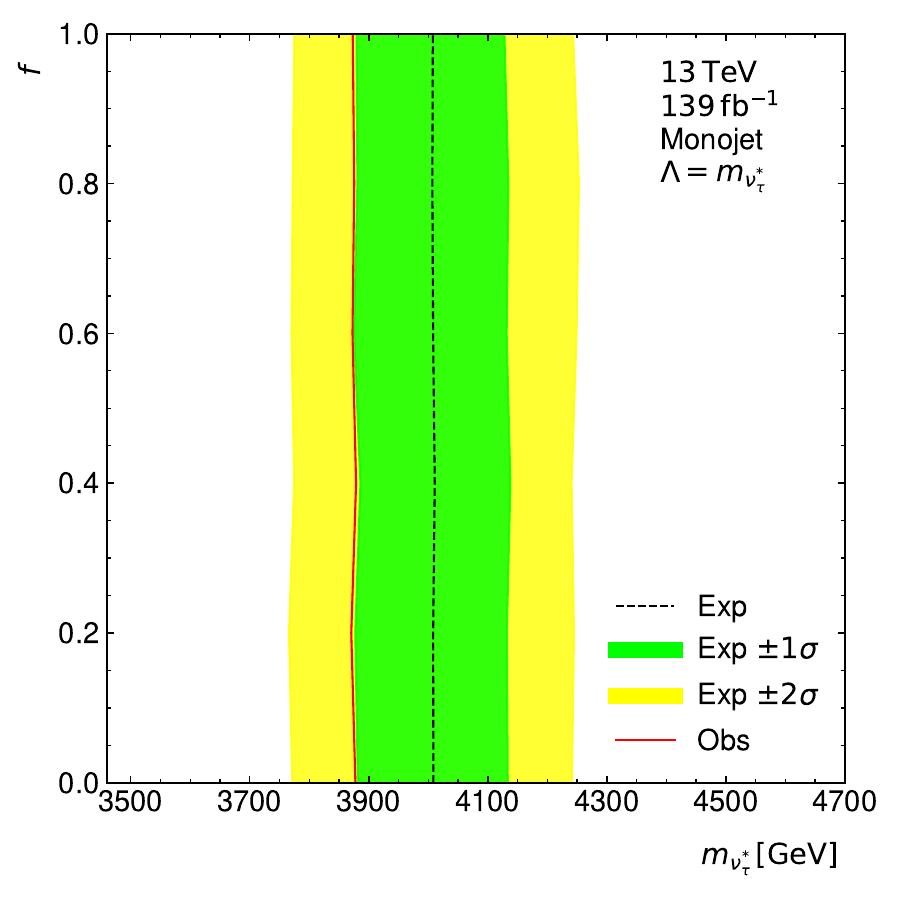}
    \caption{}
    \label{fig:atlas_contour_nuexct}
\end{subfigure}
\begin{subfigure}[b]{0.45\textwidth}
    \centering
    \includegraphics[width=\textwidth]{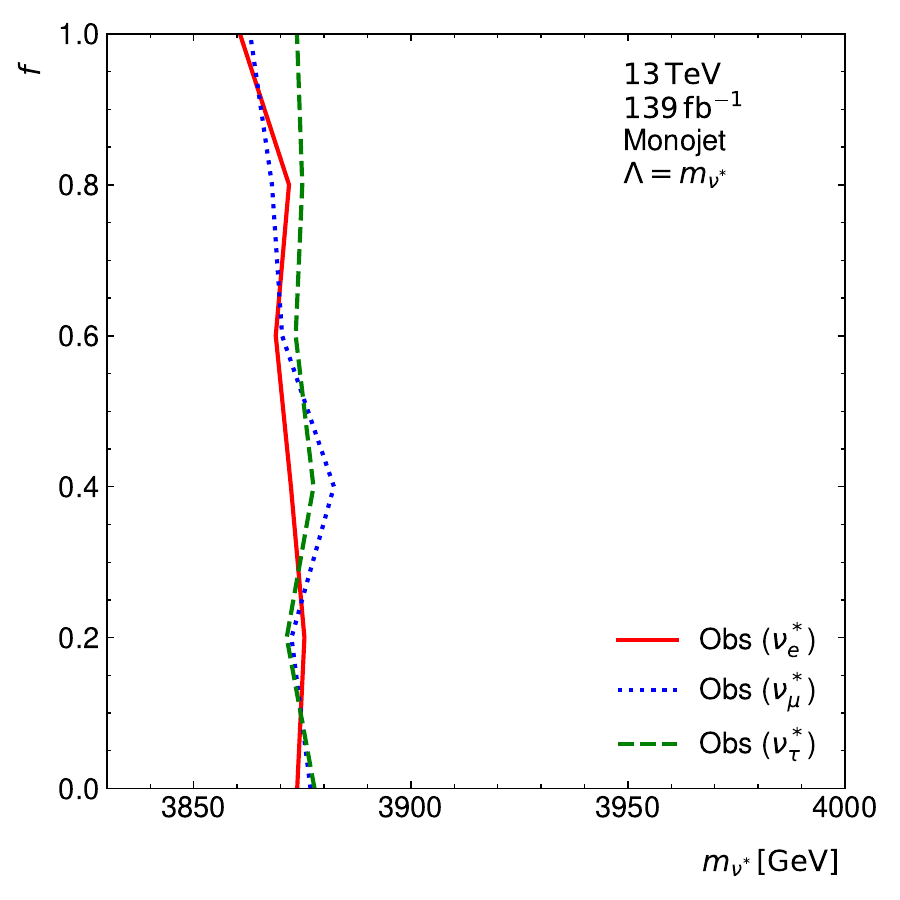}
    \caption{}
    \label{fig:atlas_contour_3nu}
\end{subfigure}
\caption{
Limits based on the ATLAS monojet search results.
(a) 95\% CL exclusion contours in the (\mnuexct, $f$) plane for the $\Lambda = \mnuexct$, $f = f'$ scenario.
The observed (expected) limit is shown in a solid red (dashed black) line.
Boundaries of the green (yellow) band display $\pm 1\sigma$ ($\pm 2\sigma$) uncertainty in the expected limit.
(b) Observed 95\% CL exclusion contours for all three \nuexc flavours in the (\mnuexc, $f$) plane for the $\Lambda = \mnuexc$, $f = f'$ scenario.
The solid red, dotted blue, and dashed green lines display the observed limits on the \nuexce, \nuexcm, and \nuexct models, respectively.
}
\label{fig:atlas_contour}
\end{figure}

Figure~\ref{fig:atlas_xs_lim} presents 95\% CL upper limits on the \nuexct production cross-section as a function of \mnuexct.
The model parameters are set in the following way: $\Lambda = \mnuexct$, $f = f' = 1$.
The observed cross-section limit is $6-8$~fb for all \mnuexct values between 2~TeV and 4.5~TeV, and it is within the $2\sigma$ band around the expected limit.
The observed mass lower limit is $\sim$3900~GeV.
Figure~\ref{fig:atlas_contour_nuexct} presents 95\% CL exclusion contours in the (\mnuexct, $f$) plane for the $\Lambda = \mnuexct$ scenario.
The observed limit is $\sim$3900~GeV for all $f$ values, and it is within the $2\sigma$ band around the expected limit.
The limits on \nuexce and \nuexcm are evaluated in analogy with the \nuexct limits.
They are shown in Figure~\ref{fig:atlas_contour_3nu} together with the \nuexct contour.
The observed mass lower limits are $\sim$3900~GeV for all \nuexc flavours, and values of $f = f'$.
The lack of $f$ dependence is expected, as the $f$-sensitive GI decays are suppressed when $\Lambda = \mnuexc$.

\section{Reinterpretation of the CMS monojet and mono-$V$ search}
\label{sec:reinterpretation_cms}

The CMS search presented in Ref.~\cite{CMS-EXO-20-004} is based on events with a high-\pt jet and large magnitude of \ptmiss, denoted as \ptmissmag in the CMS paper and this section.
It uses 13~TeV \pp collision data collected by the CMS detector~\cite{Ressegotti:2019Universe} in 2017 and 2018, corresponding to an integrated luminosity of 101~fb$^{-1}$.
It employs machine learning techniques to define two complementary final-state categories: a monojet category targeting events with narrow jets from initial-state radiation, and a mono-$V$ category optimised for events with large-radius jets consistent with hadronic decays of $W$ or $Z$ bosons.
In the SR, \ptmissmag is required to be larger than 250~GeV.
The mono-$V$ category requires the leading-\pt AK8 jet to have $\pt > 250$~GeV, $|\eta| < 2.4$, be V~tagged with the \textsc{DeepAK8} algorithm~\cite{CMS-JME-18-002}, and have a mass, modified with the soft-drop algorithm~\cite{softdrop1,PhysRevLett.100.242001}, compatible with that of a $W$ or $Z$ boson.
The mono-$V$ category is further divided into two SRs, low- and high-purity mono-$V$, depending on the \textsc{DeepAK8} score.
Events not satisfying the mono-$V$ criteria are assigned to the monojet category if the leading-\pt AK4 jet has $\pt > 100$~GeV, $|\eta| < 2.4$, and passes additional good quality criteria.
Additional SR selection criteria, such as the veto of electrons, muons, photons, hadronically decaying tau leptons, or the presence of $b$-tagged jets in the event, angular separation between \ptmiss and jets, or similarity between the default and calorimeter-based \ptmissmag value, are described in Ref.~\cite{CMS-EXO-20-004}.

The present reinterpretation study emulates the \textsc{DeepAK8} algorithm decision by applying the efficiency rates quoted in Ref.~\cite{CMS-EXO-20-004} to the leading-\pt AK8 jet in each event.
The fraction of events contributing to the high-purity mono-$V$ sample is 30\% (40\%) when the jet is truth-matched to a hadronically decaying $W$ or $Z$ boson and has $\pt = 250$~GeV ($\pt > 800$~GeV).
A linear interpolation is used to determine the fraction of events in the $\pt \in (250, 800)$~GeV region.
When the leading-\pt AK8 jet is not truth-matched to a hadronically decaying $W$ or $Z$ boson, the fraction of events contributing to the high-purity mono-$V$ sample is 0.7\%.
From events which are not added to the high-purity mono-$V$ sample, a fraction of 40\% (7\%), resp. 30\% (5\%) is assigned to the low-purity mono-$V$ sample when the leading-\pt AK8 jet is (is not) truth-matched to a hadronically decaying $W$ or $Z$ boson and has $\pt = 250$~GeV, resp. $\pt > 800$~GeV.
A linear interpolation of the efficiency rates is used in the $\pt \in (250, 800)$~GeV region.

The dominant SM backgrounds arise from $Z(\nu\nu)$+jets, $W(\ell\nu)$+jets, and $\gamma$+jets production, with smaller contributions from \ttbar, diboson, and multijet processes.
The SRs, low-purity mono-$V$, high-purity mono-$V$, and monojet, are binned in \ptmissmag.
The background in the SRs is estimated using MC simulation together with data from CRs enriched in $W(\mu\nu)$, $W(e\nu)$, $Z(\mu\mu)$, $Z(ee)$, and $\gamma$+jets events.
The CRs are binned in the magnitude of the hadronic recoil transverse momentum, which is a quantity analogous to the \ptrecmag variable defined in the ATLAS monojet analysis.
It will be denoted as \ptrecmag hereafter.
The CMS analysis provides a SR background estimate determined in a simultaneous background-only profile likelihood fit to the CRs, and it provides the data yield in each SR bin~\cite{hepdata.106115}.
It also provides the covariance matrix of nuisance parameters associated with the background estimate in the simplified likelihood approach~\cite{Collaboration:2242860}.
The signal template is obtained by applying the SR selection to the corresponding simulated signal sample for each setting of the $\Lambda$, \mnuexc, $f$, $f'$ values.
The present study uses these inputs to construct the simplified likelihood function.
Apart from the likelihood construction, the limit setting strategy is the same as described in Section~\ref{sec:reinterpretation_atlas}.
A custom script based on the iminuit package~\cite{iminuit,James:1975dr} is used to build the simplified likelihood function and perform the statistical tests.

\begin{figure}
\centering
\begin{subfigure}[b]{0.59\textwidth}
    \centering
    \includegraphics[width=\textwidth]{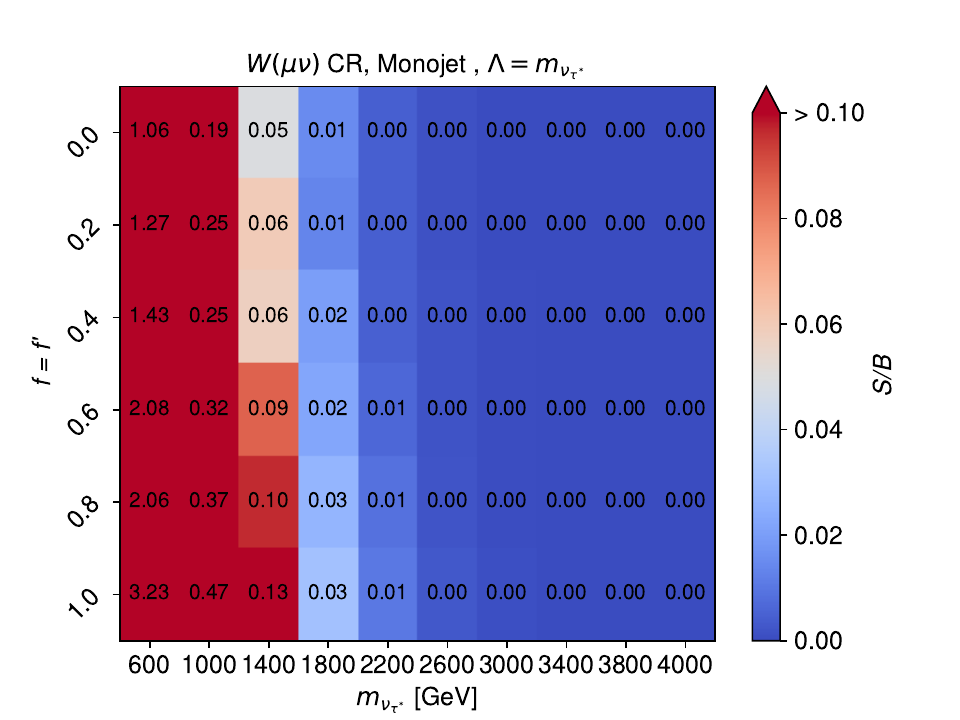}
    \caption{}
    \label{fig:cms_sig_cont}
\end{subfigure}
\begin{subfigure}[b]{0.40\textwidth}
    \centering    
    \includegraphics[width=\textwidth]{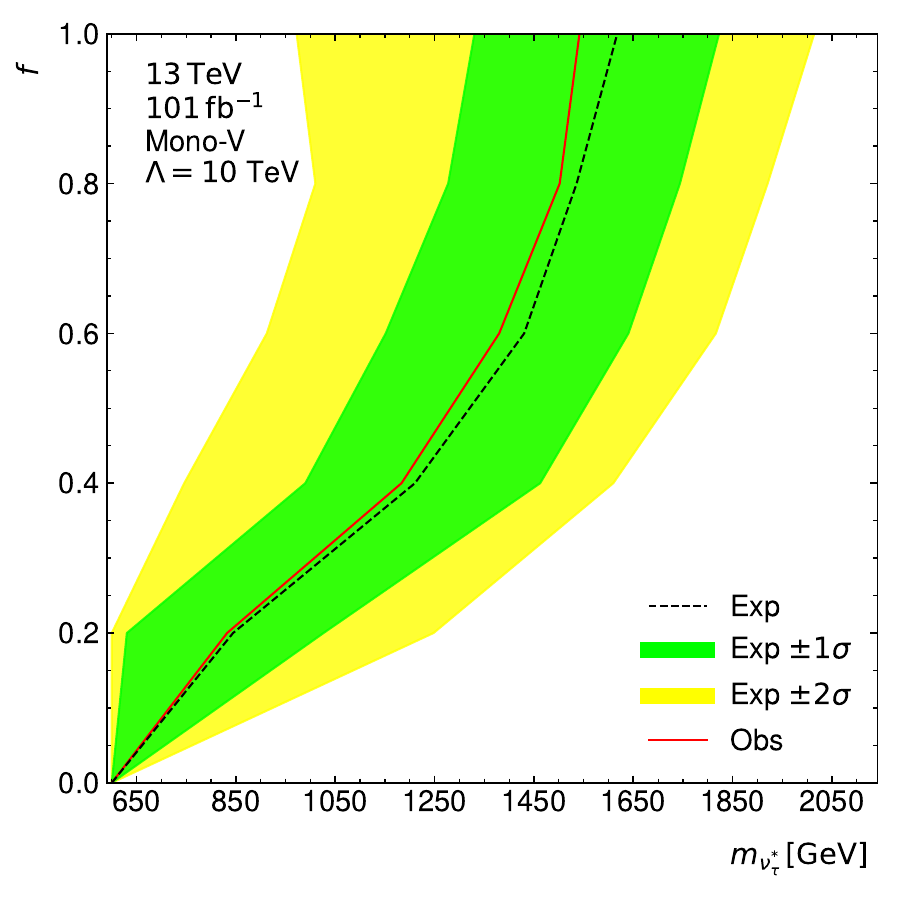}
    \caption{}
    \label{fig:cms_contour_mv_excnut}
\end{subfigure}
\caption{
(a) \sbr of \nuexct events in the high-\ptrecmag subregion of the CMS $W(\mu\nu)$ CR as a function of \mnuexct and $f$.
The \ptrecmag threshold is 600~GeV.
The $\Lambda$ parameter is set equal to \mnuexct, and $f' = f$.
The dark red colour highlights $(\mnuexct, f)$ points with signal yield higher than 10\% of the post-fit background estimate.
(b) 95\% CL exclusion contours in the (\mnuexct, $f$) plane for the $\Lambda = 10$~TeV, $f = f'$ scenario.
The observed (expected) limit is shown in a solid red (dashed black) line.
Boundaries of the green (yellow) band display $\pm 1\sigma$ ($\pm 2\sigma$) uncertainty in the expected limit.
The limits are based on the CMS mono-$V$ search results.
}
\label{fig:cms_sig_cont_contour_mv_excnut}
\end{figure}

The present study does the same check for potential signal contamination in the CMS CRs to ensure the reinterpretation's logical consistency, as described in Section~\ref{sec:reinterpretation_atlas}.
For all monojet and mono-$V$ categories, all CRs, and all \nuexc flavours, the 600~GeV threshold on \ptrecmag is found to satisfy the requirement described in Section~\ref{sec:reinterpretation_atlas}, and is used to define the high-\ptrecmag subregion of each CR.
When the $\Lambda$ parameter is set to 10~TeV, the signal contamination is negligible in the high-\ptrecmag subregion of each CR for all the monojet and mono-$V$ categories and all the considered \nuexc scenarios.
The situation is different when $\Lambda = \mnuexc$.
In the monojet category, the single muon, $W(\mu\nu)$, CR is the most affected by the \nuexct and \nuexcm signal contamination, excluding the following parameter space points from the statistical test: $\mnuexct,\ \mnuexcm \leq 1$~TeV for all $f = f'$ values, and $\mnuexct\ (\mnuexcm) \leq 1.4$~TeV for $f = f' \geq 0.8$ (0.6).
Similarly, the $W(e\nu)$ CR is the most affected by the \nuexce signal contamination, excluding $\mnuexce \leq 1$~TeV for all $f = f'$ values, and $\mnuexce \leq 1.4$~TeV for $f = f' \geq 0.6$.
Figure~\ref{fig:cms_sig_cont} displays \sbr for the \nuexct signal in the high-\ptrecmag subregion of the $W(\mu\nu)$ monojet CR.
Regarding the mono-$V$ categories, a point in the \nuexc model parameter space is excluded from the statistical test if the \sbr is above the 10\% threshold in a certain high-\ptrecmag CR subregion in either the low-purity or high-purity mono-$V$ category.
For high $f = f'$ values, masses up to 2.2~TeV are not used in the statistical test.

\begin{figure}
\centering
\begin{subfigure}[b]{0.45\textwidth}
    \centering
    \includegraphics[width=\textwidth]{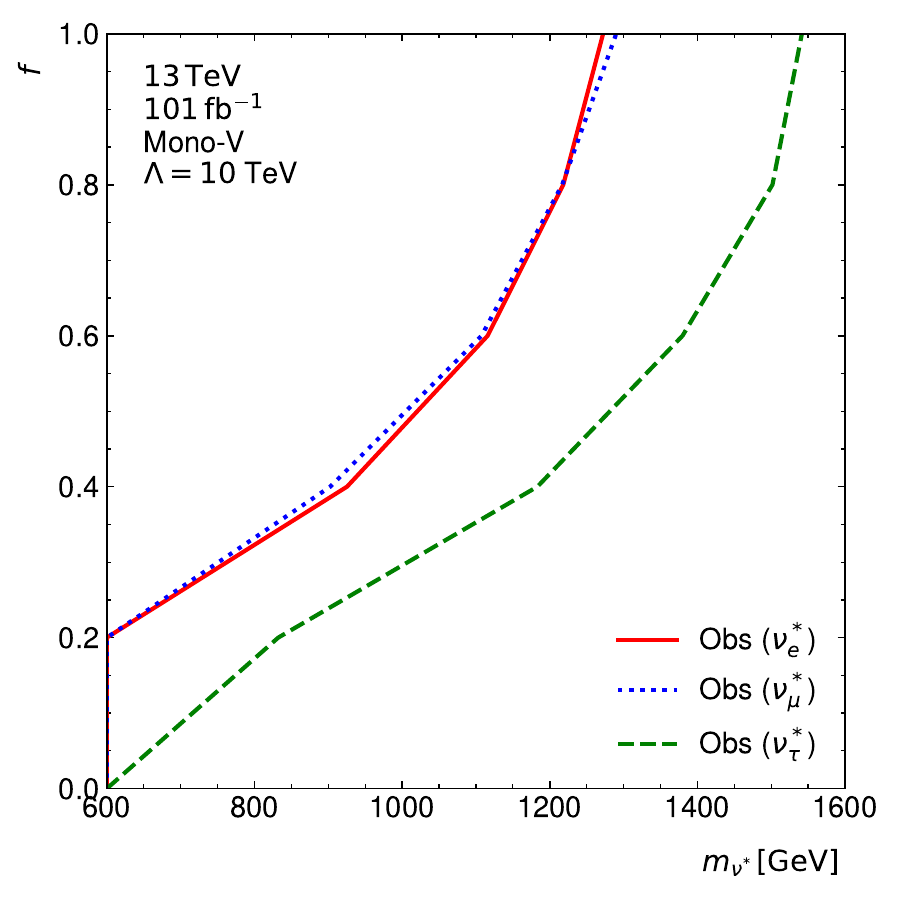}
    \caption{}
    \label{fig:cms_contour_monov}
\end{subfigure}
\begin{subfigure}[b]{0.45\textwidth}
    \centering
    \includegraphics[width=\textwidth]{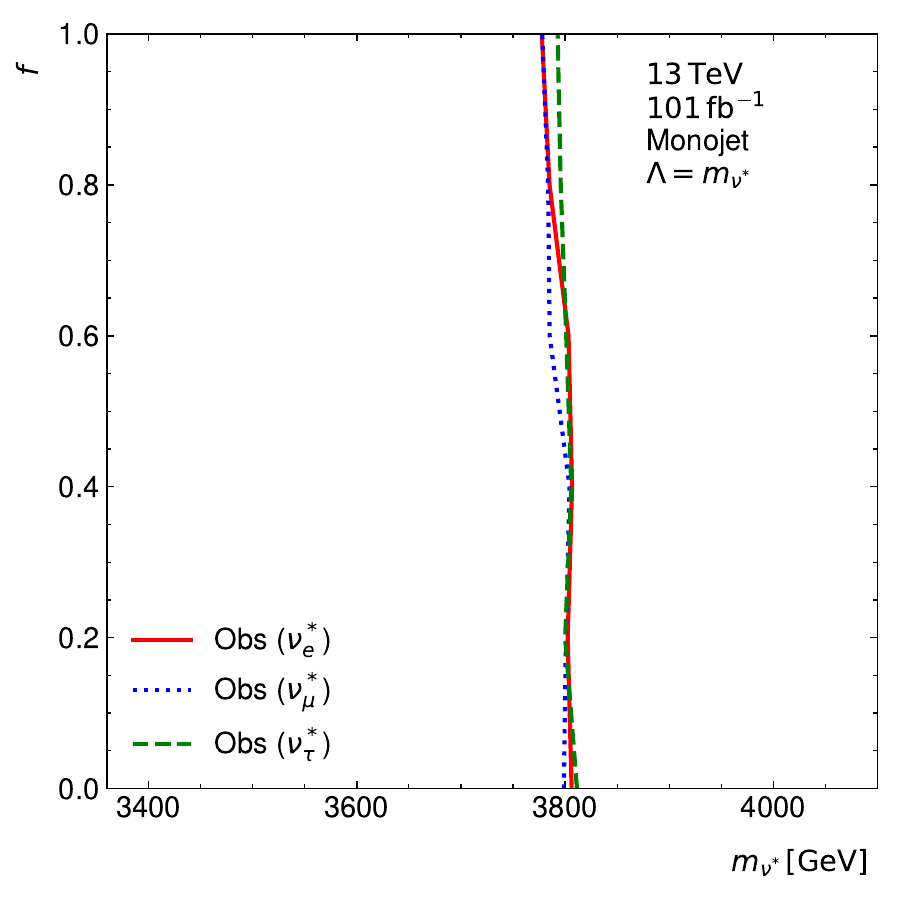}
    \caption{}
    \label{fig:cms_contour_monoj}
\end{subfigure}
\caption{
(a) Observed 95\% CL exclusion contours for all three \nuexc flavours in the (\mnuexc, $f$) plane for the $\Lambda = 10$~TeV, $f = f'$ scenario based on the CMS mono-$V$ search results.
(b) Observed 95\% CL exclusion contours for all three \nuexc flavours in the (\mnuexc, $f$) plane for the $\Lambda = \mnuexc$, $f = f'$ scenario based on the CMS monojet search results.
The solid red, dotted blue, and dashed green lines display the observed limits on the \nuexce, \nuexcm, and \nuexct models, respectively.
}
\label{fig:cms_contour}
\end{figure}

The present study uses the monojet and mono-$V$ categories separately in the statistical analysis to show their individual sensitivities to the \nuexc model.
When $\Lambda = 10$~TeV, the monojet category cannot be used to set limits on the \nuexc model for the considered mass range $\mnuexc \geq 600$~GeV, as it shows too little sensitivity to this type of signal.
The mono-$V$ SRs turn out to be sensitive to this part of the parameter space, particularly for high values of $f = f'$.
The low- and high-purity mono-$V$ SRs are used simultaneously in the statistical test to set limits on the \nuexc model.
Figure~\ref{fig:cms_contour_mv_excnut} shows the 95\% CL exclusion contours in the (\mnuexct, $f$) plane for the $f = f'$ scenario.
The \mnuexct limit reaches $\sim$1.5~TeV for $f = f' = 1$.
It is within the $1\sigma$ band around the expected limit.
Figure~\ref{fig:cms_contour_monov} shows the observed mono-$V$-data-based exclusion contours in the $(\mnuexc, f)$ parameter plane for all three \nuexc flavours, $\Lambda = 10$~TeV, and $f = f'$.
The limits on \mnuexce and \mnuexcm are less stringent than the \mnuexct ones.
In the considered parameter space region, the dominant decay channel is $\nuexc(Wl)$.
The SRs accept more of these events for the \nuexct signal than for the \nuexce and \nuexcm signals, as it is more likely to misreconstruct a hadronically decaying $\tau$ lepton as a jet than the $e$ or $\mu$ leptons.
When $\Lambda = \mnuexc$, the monojet category is more sensitive to the \nuexc signal than the mono-$V$ one.
Figure~\ref{fig:cms_contour_monoj} shows the observed monojet-data-based exclusion contours in the $(\mnuexc, f)$ parameter plane for all three \nuexc flavours, $\Lambda = \mnuexc$, and $f = f'$.
The observed mass limits are $\sim$3800~GeV with almost no dependence on $f$.
These limits are compatible with the corresponding limits based on the ATLAS monojet search results.

\section{Conclusion}
\label{sec:conclusion}

This work presents a reinterpretation of ATLAS and CMS monojet and mono-$V$ searches based on the LHC Run~2 \pp data in the context of the BSZ \nuexc model.
The most recent limit on \nuexc mass was set by ATLAS in 2012, using the 8~TeV \pp LHC Run~1 data, excluding masses below 1.6~TeV.
No comparable constraints have been reported using the significantly larger Run~2 datasets.
The present study demonstrates the potential of the ATLAS and CMS Run~2 monojet searches to exclude \nuexc with masses up to almost 4~TeV for all three \nuexc flavours at 95\% CL when $\Lambda = \mnuexc$.
This limit shows no significant dependence on the coupling parameters $f$ and $f'$,
as the monojet SRs are mostly populated by \nuexc CI decays when $\Lambda = \mnuexc$.
The CMS mono-$V$ search has the potential to constrain a complementary region of the parameter space, with lower \nuexc masses, high $\Lambda$ and high $f$, $f'$.
The mass limit can reach up to $\sim$1.5~TeV for $f = f' = 1$ and $\Lambda = 10$~TeV.

This paper is accompanied by a GitLab repository~\cite{excnu:reinterpretation} providing cross-section limits for all \nuexc flavours, $f = f'$ values and both benchmark $\Lambda$ settings. It also provides expected limits in the (\mnuexc, $f$) plane for all three \nuexc flavours and both $\Lambda$ settings, as well as more details on the high-\ptrecmag CR subregion signal contamination studies.

\FloatBarrier

\acknowledgments
This work was supported by the Grant Agency of Charles University (GA UK), project No. 284222.

\section*{Data Availability Statement}
This article has no associated data or the data will not be deposited.

\section*{Code Availability Statement}
This article has no associated code or the code will not be deposited.

\appendix

\bibliographystyle{JHEP}
\bibliography{references}

\end{document}